\documentclass{article}
\usepackage[applemac]{inputenc}
\usepackage{cite}
\usepackage{amssymb}
\usepackage{amsmath}
\usepackage{pstricks, pst-node, pst-tree}
\usepackage{fullpage}
\usepackage{relate}
\usepackage{amsthm}
\newtheorem{lemma}{Lemma}
\newtheorem{theorem}{Theorem}

\newcommand{\ceil}[1]{\left\lceil{#1}\right\rceil}

\newcommand{\Down}{\ensuremath{\textsc{Down}}}
\newcommand{\Up}{\ensuremath{\textsc{Up}}}

\newcommand{\Eq}{\ensuremath{\textsc{Eq}}}

\newcommand{\Child}{\ensuremath{\textsc{Child}}}

\newcommand{\Visit}{\ensuremath{\textsc{Visit}}}

\newcommand{\lightdepth}{\ensuremath{\mathrm{lightdepth}}}
\newcommand{\size}{\ensuremath{\mathrm{size}}}
\newcommand{\heavy}{\ensuremath{\mathrm{heavy}}}

\newcommand{\depth}{\ensuremath{\mathrm{depth}}}

\newcommand{\roots}{\ensuremath{\mathrm{root}}}
\newcommand{\parent}{\ensuremath{\mathrm{parent}}}

\newcommand{\child}{\ensuremath{\mathrm{child}}}
\newcommand{\lab}{\ensuremath{\mathrm{label}}}
\newcommand{\path}{\ensuremath{\mathrm{path}}}

\title{Matching Subsequences in Trees\footnote{An extended abstract of this paper appeared in Proceedings of the 6th Italian Conference on Algorithms and Complexity, 2006.}}

\author{Philip Bille\thanks{The IT University
of Copenhagen, Rued
Langgaards Vej 7, DK-2300 Copenhagen S, Denmark. Email: {\tt
beetle@itu.dk}. This work is part of the DSSCV project
supported by the IST Programme of the European Union
(IST-2001-35443).} \and Inge Li G{\o}rtz\thanks{Technical University of Denmark, Department of Informatics and Mathematical Modelling, Building 322, DK-2800 Kongens Lyngby,
Denmark. Email: {\tt ilg@imm.dtu.dk}. This work was performed while the author was a PhD student at the IT University of Copenhagen.}}

\date{\today}

\begin{document}
\maketitle
\begin{abstract}
Given two rooted, labeled trees $P$ and $T$ the tree path
subsequence problem is to determine which paths in $P$ are
subsequences of which paths in $T$. Here a path begins at the root and
ends at a leaf. In this paper we propose this problem as a useful
query primitive for XML data, and provide new algorithms improving
the previously best known time and space bounds.
\end{abstract}

\section{Introduction}
We say that a tree is \emph{labeled} if each node is assigned a
character from an alphabet $\Sigma$. Given two sequences of
labeled nodes $p$ and $t$, we say that $p$ is a \emph{subsequence}
of $t$, denoted $p \sqsubseteq t$, if $p$ can be obtained by
removing nodes from $t$. Given two rooted, labeled trees $P$ and
$T$ the \emph{tree path subsequence problem} (TPS) is to determine
which paths in $P$ are subsequences of which paths in $T$. Here a path
begins at the root and ends at a leaf. That is, for each path $p$ 
in $P$ we must report all paths $t$ in $T$ such that $p \sqsubseteq t$.

This problem was introduced by Chen~\cite{Chen2000} who gave an
algorithm using $O(\min(l_{P}n_{T} + n_P, n_{P}l_{T}+n_T))$ time and
$O(l_{P}d_{T} + n_P + n_T)$ space. Here, $n_{S}$, $l_{S}$, and $d_{S}$ denotes
the number of nodes, number of leaves, and depth, respectively, of
a tree $S$. Note that in the worst-case this is quadratic time and
space. In this paper we present improved algorithms giving the following result:
\begin{theorem}\label{main}
For trees $P$ and $T$ the tree path subsequence problem can be 
solved in $O(n_{P} + n_{T})$ space with the following running times:
\begin{equation*}
\min
\begin{cases}
    O(l_{P}n_{T} + n_P) , \\
     O(n_{P}l_{T}+n_T) , \\
     O(\frac{n_{P}n_{T}}{\log n_{T}}+ n_T + n_P \log n_P).
\end{cases}
\end{equation*}
\end{theorem}
The first two bounds in Theorem~\ref{main} match the previous time bounds while improving the space to linear. This is achieved using a algorithm that resembles the algorithm of Chen~\cite{Chen2000}. At a high level, the algorithms are essentially identical and therefore the bounds should be regarded as an improved analysis of Chen's algorithm. The latter bound is obtained by using an entirely new algorithm that improves the worst-case quadratic time. Specifically, whenever $\log n_{P} = O(n_{T}/ \log n_{T})$ the running time is improved by a logarithmic factor. Note that -- in the worst-case -- the number of pairs consisting of a path from $P$ and a path $T$ is $\Omega(n_{P}n_{T})$, and therefore we need at least as many bits to report the solution to TPS. Hence, on a RAM with logarithmic word size our worst-case bound is optimal. Most importantly, all our algorithms use linear space. For practical applications this will likely make it possible to solve TPS on large trees and improve running time since more of the computation can be kept in main memory.

\subsection{Applications} We propose TPS as a useful query
primitive for XML data. The key idea is that an XML document $D$
may be viewed as a rooted, labeled tree (different nodes can be assigned the same label).
\begin{figure}[t]
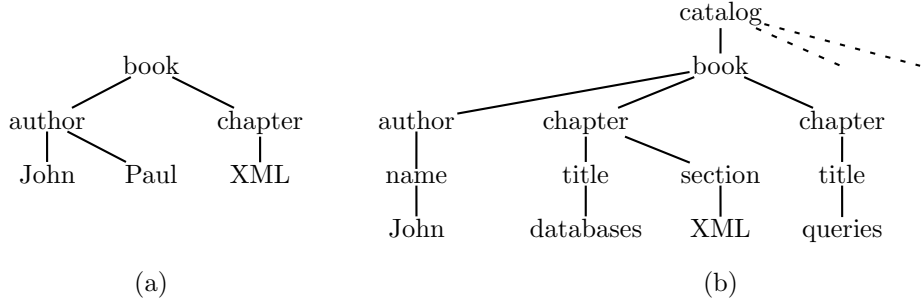

\begin{center}
  \begin{psmatrix}[colsep=0.5cm,rowsep=0.3cm,labelsep=1pt, nodesep=1pt]
  &&&&&&&&[name=cat] catalog \\
  &&[name=book] book &&&&&&[name=book2] book & [name=book3] & [name=book4]\\
  & [name=author] author &&[name=chapter] chapter&&
  [name=author2] author && [name=chapter2] chapter && [name=chapter3] chapter \\
  & [name=john] John & [name=paul] Paul & [name=xml] XML &&
  [name=name] name && [name=title] title  & [name=section] section & [name=title2] title \\
  &&&&&  [name=john2] John && [name=DB] databases & [name=xml2] XML & [name=queries] queries \\
  && (a) &&&&&& (b) 
  \ncline{author}{book} \ncline{chapter}{book}
  \ncline{john}{author} \ncline{paul}{author} \ncline{xml}{chapter}
  \ncline{cat}{book2}\psset{linestyle=dashed,dash=2pt 4pt}\ncline{cat}{book3} \ncline{cat}{book4}
  \psset{linestyle=solid}
  \ncline{book2}{author2}\ncline{book2}{chapter2}\ncline{book2}{chapter3}
  \ncline{author2}{name}\ncline{chapter2}{title}\ncline{chapter2}{section}\ncline{chapter3}{title2}
  \ncline{name}{john2}\ncline{title}{DB}\ncline{section}{xml2}\ncline{title2}{queries}
  \end{psmatrix}
   \caption{(a) The trie of queries 1,2,3, or the tree for query 4. (b)
   A fragment of a catalog of books.}
  \label{xmlexample}
  \end{center}
\end{figure}
For example, suppose that we want to maintain a catalog of books
for a bookstore. A fragment of a possible XML tree, denoted $D$,
corresponding to the catalog is shown in Fig.~\ref{xmlexample}(b).
In addition to supporting full-text queries, such as find all
documents containing the word ``John'', we can also use the tree
structure of the catalog to ask more specific queries, such as the
following examples:
\begin{enumerate}
  \item Find all books written by John,
  \item find all books written by Paul,
  \item find all books with a chapter that has something to do with XML, or
  \item find all books written by John and Paul with a chapter that has something to do with XML.
\end{enumerate}
The queries 1,2, and 3 correspond to a \emph{path query} on $D$,
that is, compute which paths in $D$ that contains a specific path
as a subsequence. For instance, computing the paths in $D$ that
contain the path of three nodes labeled ``book'', ``chapter'', and
``XML'', respectively, effectively answers query 3. Most XML-query
languages, such as XPath \cite{CD1999}, support such queries.

Using a depth-first traversal of $D$ a path query can be solved in linear time.
More precisely, if $q$ is a path consisting of $n_{q}$ nodes, answering the path query on $D$ takes $O(n_{q} + n_{D})$ time. Hence, if we are given path queries $q_{1}, \ldots, q_{k}$ we can
answer them in $O(n_{q_{1}} + \cdots + n_{q_{k}} + kn_{D})$ time. However, we can do better by constructing the \emph{trie}, $Q$, of $q_{1}, \ldots, q_{k}$. The trie $Q$ has labels on the nodes and is constructed such there is one node for every common prefix of $q_{1}, \ldots, q_{k}$. 
Answering all path queries now correspond to solving TPS on $Q$ and $D$. As an example the queries 1,2, and 3 form the trie shown in Fig.~\ref{xmlexample}(a). As $l_Q \leq k$, Theorem~\ref{main} gives us an algorithm with running time 
\begin{equation}
\label{bound}
O\left(n_{q_{1}} + \cdots + n_{q_{k}} + \min\left(kn_D + n_Q, n_Ql_D + n_D, \frac{n_Qn_D}{\log n_D} + n_D + n_Q\log n_Q\right)\right). 
\end{equation}
Since $n_Q \leq n_{q_{1}} + \cdots + n_{q_{k}}$ this is at least as good as answering the queries individually and better in many cases. If many paths share a prefix, i.e., queries 1 and 2 share "book" and "author", 
the size of $n_Q$ can be much smaller than $n_{q_{1}} + \cdots + n_{q_{k}}$. Using our solution to TPS we can efficiently take advantage of this situation since the latter two terms in \eqref{bound} depend on $n_Q$ and not on $n_{q_{1}} + \cdots + n_{q_{k}}$.  

Next consider query 4. This query cannot be answered by solving a
TPS problem but is an instance of the \emph{tree inclusion
problem} (TI). Here we want to decide if $P$ is \emph{included} in
$T$, that is, if $P$ can be obtained from $T$ by \emph{deleting}
nodes of $T$. Deleting a node $y$ in $T$ means making the children
of $y$ children of the parent of $y$ and then removing $y$. It is
straightforward to check that we can answer query 4 by deciding if
the tree in Fig.~\ref{xmlexample}(a) can be included in the tree
in Fig.~\ref{xmlexample}(b).

Recently, TI has been recognized as an important XML query
primitive and has recieved considerable attention, see e.g.,
\cite{SM2002, YLH2003,YLH2004, ZADR03, SN2000, TRS2002}.
Unfortunately, TI is NP-complete in general 
\cite{KM1995} and therefore the existing algorithms are based on
heuristics. Observe that a necessary condition for $P$ to included
in $T$ is that all paths in $P$ are subsequences of paths in $T$.
Hence, we can use TPS to quickly identify trees or parts of trees that cannot be
included $T$. We believe that in this way TPS can be used as an
effective "filter" for many tree inclusion problems that occur in
practice.

We note that for \emph{ordered} trees, that is, a left-to-right
ordering among siblings is given, the tree inclusion problem can
be solved in polynomial time~\cite{KM1995, BG2005}. In this case
deleting a node $y$ inserts the children of $y$ in the place of in
the left-to-right order among the siblings of $y$.

\subsection{Technical Overview} Given two strings (or labeled
paths) $a$ and $b$, it is straightforward to determine if $a$ is a
subsequence of $b$ by scanning the character from left to right in
$b$. This uses $O(|a| + |b|)$ time. We can solve TPS by applying
this algorithm to each of the pair of paths in $P$ and $T$,
however, this may use as much as $O(n_Pn_T(n_P + n_T))$ time.
Alternatively, Baeza-Yates \cite{BaezaYates1991} showed how to
preprocess $b$ in $O(|b|\log |b|)$ time such that testing whether
$a$ is a subsequence of $b$ can be done in $O(|a|\log |b|)$ time.
Using this data structure on each path in $T$ we can solve the TPS
problem, however, this may take as much as $O(n^2_{T}\log n_{T} +
n_{P}^{2}\log n_{T})$. Hence, none of the availiable subsequence algorithms
on strings provide an immediate efficient solution to TPS. 

Inspired by the work of Chen~\cite{Chen2000} we take another
approach. We provide a framework for solving TPS. The main
idea is to traverse $T$ while maintaining a subset of nodes in
$P$, called the \emph{state}. When reaching a leaf $z$ in $T$ the
state represents the paths in $P$ that are subsequences of the
path from the root to $z$. At each step the state is updated using a simple
procedure processing a subset of nodes. The result of
Theorem~\ref{main} is obtained by taking the best of two
algorithms based on our framework: The first one uses a simple
data structure to maintain the state. This leads to an algorithm
using $O(\min(l_{P}n_{T} + n_P, n_{P}l_{T}+n_T))$ time. At a high level this
algorithm resembles the algorithm of Chen~\cite{Chen2000} and
achieves the same running time. However, we improve the analysis
of the algorithm and show a space bound of  $O(n_P +
n_T)$. This should be compared to the worst-case quadratic space
bound of $O(l_{P}d_{T} + n_P + n_T)$ given by Chen~\cite{Chen2000}. Our second
algorithm takes a different approach combining several 
techniques. Starting with a simple quadratic time and space
algorithm, we show how to reduce the space to $O(n_P \log n_T)$
using a decomposition of $T$ into disjoint paths. We then divide
$P$ into small subtrees of logarithmic size called \emph{micro
trees}. The micro trees are then  preprocessed such that subsets of
nodes in a micro tree can be maintained in constant time and
space. Intuitively, this leads to a logarithmic improvement of the time and space bounds.



\subsection{Notation and Definitions} In this section we define
the notation and definitions we will use throughout the paper. For
a graph $G$ we denote the set of nodes and edges by $V(G)$ and
$E(G)$, respectively. Let $T$ be a rooted tree. The root of $T$ is
denoted by $\roots(T)$. The \emph{size} of $T$, denoted by $n_T$,
is $|V(T)|$. The \emph{depth} of a node $y\in V(T)$, $\depth(y)$,
is the number of edges on the path from $y$ to $\roots(T)$ and the
depth of $T$, denoted $d_T$, is the maximum depth of any node in
$T$. The parent of $y$ is denoted $\parent(y)$. A node with no
children is a leaf and otherwise it is an internal node. The
number of leaves in $T$ is denoted $l_T$. Let $T(y)$ denote the
subtree of $T$ rooted at a node $y \in V(T)$. If $z\in V(T(y))$
then $y$ is an ancestor of $z$ and if $z\in V(T(y))\backslash
\{y\}$ then $y$ is a proper ancestor of $z$. If $y$ is a (proper)
ancestor of $z$ then $z$ is a (proper) descendant of $y$. We say
that $T$ is \emph{labeled} if each node $y$ is assigned a
character, denoted $\lab(y)$, from an alphabet $\Sigma$. The path
from $y$ to $\roots(T)$, of nodes $\roots(T) = y_1, \ldots, y_k =
y$ is denoted $\path(y)$. Hence, we can formally state TPS as
follows: Given two rooted tree $P$ and $T$ with leaves $x_1,
\ldots, x_r$ and $y_1, \ldots, y_s$, respectively, determine all
pairs $(i,j)$ such that $\path(x_{i}) \sqsubseteq \path(y_{j})$.
For simplicity we will assume that leaves in $P$ and $T$ are
always numbered as above and we identify each of the paths by the
number of the corresponding leaf.

Throughout the paper we assume a unit-cost RAM model of computation
with word size $\Theta(\log n_T)$ and a standard instruction set including bitwise boolean operations, shifts, addition and multiplication. All space complexities refer to the number of words used by the algorithm.

\section{A Framework for solving TPS}
In this section we present a simple general
algorithm for the tree path subsequence problem. The key
ingredient in our algorithm is the following procedure. For any $X
\subseteq V(P)$ and $y \in V(T)$ define:
\begin{relate}
\item[$\Down(X,y)$:] Return the set
    $\Child(\{x \in X \mid \lab(x) = \lab(y)\})
    \cup \{x \in X \mid \lab(x) \neq \lab(y)\}$.
\end{relate}
The notation $\Child(X)$ denotes the set of children of $X$. 
Hence, $\Down(X,y)$ is the set consisting of nodes in $X$ with a
different label than $y$ and the children of the nodes $X$ with
the same label as $y$. We will now show how to solve TPS using
this procedure.

First assign a unique number in the range $\{1,\ldots,l_{P}\}$ to
each leaf in $P$. Then, for each $i$, $1\leq i \leq l_{P}$, add a
\emph{pseudo-leaf} $\bot_{i}$ as the single child of the $i$th
leaf. All pseudo-leaves are assigned a special label $\beta
\not\in \Sigma$. The algorithm traverses $T$ in a depth first
order and computes at each node $y$ the set $X_{y}$. We call this
set the \emph{state} at $y$. Initially, the state consists of
$\{\roots(P)\}$. For $z \in \child(y)$, the state $X_z$ can be
computed from state $X_y$ as 
\begin{equation*}
X_{z} =   \Down(X_{y}, z).
\end{equation*}
If $z$ is a leaf we report the number of each pseudo-leaf in $X_z$
as the paths in $P$ that are subsequences of $\path(z)$. See
Figure~\ref{fig:framework}  for an example. To show
the correctness of this approach we need the following lemma.

\begin{figure}[tb]
\begin{center}
\begin{psmatrix}[colsep=0.6cm,rowsep=0.3cm,labelsep=3pt]
  && \cnode{.2}{rootp}\rput(0,0){$a$}\rput(0,.4){$\roots(P)$} &&&&&&
  \cnode{.2}{roott}\rput(0,0){$a$}\rput(0,.4){$\roots(T)$}
  \\
  &\cnode{.2}{v1}\rput(0,0){$c$}\rput(.5,0){$x_1$} & &
  \cnode{.2}{v2}\rput(0,0){$b$}\rput(.5,0){$x_2$}
  &&&&& \cnode{.2}{w1}\rput(0,0){$c$}\rput(.4,0){$1$}\\
 &\cnode{.2}{v3}\rput(0,0){$a$}\rput(.5,0){$x_3$} & &
  [name=b2]\psframebox{}\rput(.3,0){$\bot_2$}
  &&&&\cnode{.2}{w2}\rput(0,0){$a$}\rput(.4,0){$2$} & &
  \cnode{.2}{w3}\rput(0,0){$b$}\rput(.4,0){$4$}
  \\
  &[name=b1] \psframebox{}\rput(.3,0){$\bot_1$}
  &&&&&&\cnode{.2}{w4}\rput(0,0){$b$}\rput(.4,0){$3$} & &
  \cnode{.2}{w5}\rput(0,0){$b$}\rput(.4,0){$5$}
  \\
  && $P$ &&&&&& $T$

  \ncline{rootp}{v1}\ncline{rootp}{v2}
  \ncline{v1}{v3}\ncline{v2}{b2}\ncline{v3}{b1}
  \ncline{roott}{w1}
  \ncline{w1}{w3}\ncline{w1}{w2}\ncline{w2}{w4}
    \ncline{w3}{w5}
  \end{psmatrix}
  \caption{The letters inside the nodes are the labels, and the
  identifier of each node is written outside the node. Initially we have $X=\{\roots(P)\}$.
  Since $\lab(\roots(P))=a=\lab(\roots(T))$ we replace $\roots(P)$ with is children and
  get
  $X_{\roots(T)}=\{x_1,x_2\}$. Since
  $\lab(1)=\lab(x_1)\neq\lab(x_2)$ we get
  $X_1=\{x_3,x_2\}$. Continuing this way we get
  $X_2=\{\bot_1,x_2\}$, $X_3=\{\bot_1,\bot_2\}$, $X_4=\{x_3,\bot_2\}$,
  and $X_5=\{x_3,\bot_2\}$. The nodes $3$ and $5$ are leaves of $T$ and we thus
  report paths $1$ and $2$ after computing $X_3$ and path $2$ after computing
  $X_5$.}\label{fig:framework}
  \end{center}
\end{figure}
\begin{lemma}\label{lem:invariant}
For any node $y \in V(T)$ the state $X_{y}$ satisfies the
following property: 
$$x \in X_{y} \Rightarrow \path(\parent(x)) \sqsubseteq \path(y)\;.$$
\end{lemma}
\begin{proof}
By induction on the number of iterations of the procedure.
Initially, $X = \{\roots(P)\}$ satisfies the property
 since $\roots(P)$ has no parent. Suppose that
$X_{y}$ is the current state and $z\in \child(y)$ is the next node
in the depth first traversal of $T$. By the induction hypothesis
$X_{y}$ satisfies the property, that is, for
any $x \in X_{y}$, $\path(\parent(x)) \sqsubseteq \path(y))$. Then, 
\begin{equation*}
X_{z} = \Down(X_{y},z) =  \Child(\{x \in X_{y} \mid \lab(x) =
\lab(z)\}) \cup \{x \in X_{y} \mid \lab(x) \neq \lab(z)\}\;.
\end{equation*}
Let $x$ be a node in $X_y$. There are two cases. If $\lab(x) =
\lab(z)$ then $\path(x) \sqsubseteq \path(z)$ since
$\path(\parent(x)) \sqsubseteq \path(y)$. Hence, for any child
$x'$ of $x$ we have $\path(\parent(x')) \sqsubseteq \path(z)$. On
the other hand, if $\lab(x) \neq \lab(z)$ then $x \in X_z$. Since
$y=\parent(z)$ we have $\path(y) \sqsubseteq \path(z)$, and hence
$\path(\parent(x)) \sqsubseteq \path(y) \sqsubseteq \path(z)$.
\end{proof}
By the above lemma all paths reported at a leaf $z
\in V(T)$ are subsequences of $\path(z)$. The following lemma shows that the paths reported at a leaf $z \in
V(T)$ are \emph{exactly} the paths in $P$ that are subsequences of $\path(z)$.
\begin{lemma}\label{lem:correct}
Let $z$ be a leaf in $T$ and let $\bot_i$ be a pseudo-leaf in $P$.
Then,
\begin{equation*}
\bot_i \in X_{z} \Leftrightarrow
\path(\parent(\bot_i)) \sqsubseteq \path(z)\;.
\end{equation*}
\end{lemma}
\begin{proof}
It follows immediately from Lemma~\ref{lem:invariant} that $\bot_i
\in X_{z} \Rightarrow \path(\parent(\bot_i)) \sqsubseteq
\path(z)$. It remains to show that $\path(\parent(\bot_i))
\sqsubseteq \path(z) \Rightarrow \bot_i \in X_{z}$. Let
$\path(z)=z_1,\ldots,z_k$, where $z_1=\roots(T)$ and $z_k=z$, and
let $\path(\parent(\bot_i))=y_1,\ldots,y_\ell$, where
$y_1=\roots(P)$ and $y_\ell=\parent(\bot_i)$. Since
$\path(\parent(\bot_i)) \sqsubseteq \path(z)$ there are nodes
$z_{j_i}=y_i$ for $1\leq i\leq k$, such that (i) $j_i< j_{i+1}$
and (ii) there exists no node $z_j$ with $\lab(z_j)=\lab(y_i)$,
where $j_{i-1}<j<j_i$. Initially, $X=\{\roots(P)\}$. We have
$\roots(P) \in X_{z_j}$ for all $j<j_{1}$, since $z_{j_1}$ is the
first node on $\path(z)$ with label $\lab(\roots(P))$. When we get
to $z_{j_1}$, $\roots(P)$ is removed from the state and $y_2$ is
inserted. Similarly, $y_i$ is in all states $X_{z_j}$ for $j_{i-1}
\leq j <j_i$. It follows that $\bot_i$ is in all states $X_{z_j}$
where $j\geq j_\ell$ and thus $\bot_i \in X_{z_k}=X_z$.
\end{proof}
The next lemma can be used to give an upper bound on the number of
nodes in a state.
\begin{lemma}\label{lem:statesize}
For any $y\in V(T)$ the state $X_y$ has the following property:
Let $x \in X_y$. Then no ancestor of $x$ is in $X_y$.
\end{lemma}
\begin{proof}
 By induction on the length of $\path(y)$.
Initially, the state only contains $\roots(P)$. Let $z$ be the
parent of $y$, and thus $X_y$ is computed from $X_z$. First we
note that for all nodes $x\in X_y$ either $x\in X_z$ or
$\parent(x)\in X_z$. If $x\in X_z$ it follows from the induction
hypothesis that no ancestor of $x$ is in $X_z$, and thus no
ancestors of $x$ can be in $X_y$. If $\parent(x)\in X_z$ then due
to the definition of \Down\ we must have $\lab(x)=\lab(y)$. It
follows from the definition of \Down\ that $\parent(x)\not \in
X_y$.
\end{proof}
It follows from Lemma~\ref{lem:statesize} that $|X_y|\leq l_P$ for
any $y\in V(T)$. If we store the
state in an unordered linked list each step of the depth-first
traversal takes time $O(l_{P})$ giving a total $O(l_{P}n_{T} +n_{P})$
time algorithm.
Since each state is of size at most $l_P$ the space used is
$O(n_{P} + l_P n_{T})$. In the following sections we show how to improve these bounds.

\section{A Simple Algorithm}
In this section we consider a simple
implementation of the above algorithm, which has running time
$O\left(\min(l_Pn_T +n_P, n_Pl_T + n_T)\right)$ and uses $O(n_P + n_T)$ space. We assume
that the size of the alphabet is $n_T + n_P$ and each character in
$\Sigma$ is represented by an integer in the range $\{1,\ldots,n_T
+ n_P\}$. If this is not the case we can sort all characters in
$V(P) \cup V(T)$ and replace each label by its rank in the sorted
order. This does not change the solution to the problem, and
assuming at least a logarithmic number of leaves in both trees it
does not affect the running time. To get the space usage down to
linear we will avoid saving all states. For this purpose we
introduce the procedure \Up, which reconstructs the state $X_z$
from the state $X_y$, where $z=\parent(y)$. We can thus save space
as we only need to save the current state.

We use the following data structure to represent the current state $X_y$: A
\emph{node dictionary} consists of two dictionaries denoted
$X^{c}$ and $X^{p}$.  The dictionary $X^c$ represents the node set
corresponding to $X_y$, and the dictionary $X^p$ represents the
node set corresponding to the set $\{x \in X_z \mid x \not\in X_y
\text{ and } z \text{ is an ancestor of } y\}$.  That is, $X^c$
represents the nodes in the current state, and $X^p$ represents
the nodes that is in a state $X_z$, where $z$ is an ancestor of
$y$ in $T$, but not in $X_y$. We will use $X^p$ to reconstruct
previous states. The dictionary $X^c$ is indexed by $\Sigma$ and
$X^p$ is indexed by $V(T)$.
The subsets stored at each entry are represented by doubly-linked
lists. Furthermore, each node in $X^{c}$ maintains a pointer to
its parent in $X^{p}$ and each node $x'$ in $X^{p}$ stores a
linked list of pointers to its children in $X^{p}$.
With this representation the total size of the node dictionary is
$O(n_{P}+n_T)$.

%
Next we show how to solve the tree path subsequence problem in our
framework using the node dictionary representation. For
simplicity, we add a node $\top$ to $P$ as a the parent of
$\roots(P)$. Initially, the $X^{p}$ represents $\top$ and $X^{c}$
represents $\roots(P)$. The $\Down$ and $\Up$ procedures are
implemented as follows:
\begin{relate}
\item[$\Down((X^{p}, X^{c}), y)$:]\begin{enumerate}
\item Set
$X:=X^c[\lab(y)]$ and $X^c[\lab(y)]:=\emptyset$.

\item For each $x\in X$ do:
\begin{enumerate}
\item Set 
$X^p[y] := X^p[y] \cup \{x\}$. 
\item For each $x' \in \child(x)$ do:
\begin{enumerate}
  \item Set $X^c[\lab(x')] := X^c[\lab(x')] \cup \{x\}$.
  \item Create pointers between $x'$ and $x$.
\end{enumerate}
\end{enumerate}
\item Return $(X^{p}, X^{c})$.
\end{enumerate}
\item[$\Up((X^{p}, X^{c}), y)$:]
\begin{enumerate}
\item Set $X:=X^p[y]$ and $X^p[y]:=\emptyset$.

\item For each $x \in X$ do:
\begin{enumerate}
\item Set
$X^c[\lab(x)] := X^c[\lab(x)] \cup \{x\}$.
\item For each $x' \in \child(x)$ do:
\begin{enumerate}
\item Remove pointers between $x'$ and $x$.
\item Set $X^c[\lab(x')] := X^c[\lab(x')] \setminus \{x'\}$.
\end{enumerate}
\end{enumerate}
\item Return $(X^{p}, X^{c})$.
\end{enumerate}
\end{relate}
The next lemma shows that 
\Up\ correctly reconstructs
the former state.
\begin{lemma}\label{lem:updown}
Let $X_z=(X^c,X^p)$ be a state computed at a node $z\in V(T)$, and
let $y$ be a child of $z$. Then, 
\begin{equation*}
X_z=\Up(\Down(X_z,y),y)\;.
\end{equation*}
\end{lemma}
\begin{proof}
Let $(X^c_1,X^p_1)= \Down(X_z,y)$ and $(X^c_2,X^p_2)=
\Up((X^c_1,X^p_1),y)$.
We will first show that $x \in X_z \Rightarrow x \in
\Up(\Down(X_z,y),y)$.

Let $x$ be a node in $X^c$. There are two cases. If $x \in
X^c[\lab(y)]$, then it follows  from the implementation of \Down\
that $x \in X^p_1[y]$. By the implementation of \Up, $x \in
X^p_1[y]$ implies $x \in X^c_2$. If $x \not\in X^c[\lab(y)]$ then
$x\in X^c_1$. We need to show $\parent(x) \not\in X^p_1[y]$. This
will imply $x\in X^c_2$, since the only nodes removed from $X^c_1$
when computing $X^c_2$ are the nodes with a parent in $X^p_1[y]$.
Since $y$ is unique it follows from the implementation of \Down\
that $\parent(x)\in X^p_1$ implies $x\in X^c[\lab(y)]$.

Let $x$ be a node in $X^p$. Since $y$ is unique we have $x \in
X^p[y']$ for some $y'\neq y$.  It follows immediately from the
implementation of \Up\ and \Down\ that
$X^p[y']=X^p_1[y']=X^p_2[y']$, when $y'\neq y$, and thus
$X^p=X^p_2$.

We will now show $x \in \Up(\Down(X_z,y),y) \Rightarrow x \in
X_z$.
Let $x$ be a node in $X^c_2$.  There are two cases. If $x \not \in
X^c_1$ then it follows from the implementation of \Up\ that $x \in
X^p_1[y]$. By the implementation of \Down, $x \in X^p_1[y]$
implies $x \in X^c[\lab(y)]$, i.e., $x\in X^c$. If $x\in X_1^c$
then by the implementation of \Up, $x \in X^c_2$ implies
$\parent(x) \not\in x^p_1[y]$. It follows from the implementation
of \Down\ that $x\in X^c$. Finally, let $x$ be a node in $X^p_2$.
As argued above $X^p=X^p_2$, and thus $x \in X^p$.
\end{proof}
From the current state $X_{y}=(X^c,X^p)$ the next state $X_{z}$ is
computed as follows:
\begin{equation*}
X_{z} =
\begin{cases}
   \Down(X_{y}, z)   & \text{if $y = \parent(z)$}, \\
   \Up(X_{y},y)  & \text{if $z = \parent(y)$}.
\end{cases}
\end{equation*}
The correctness of the algorithm follows from
Lemma~\ref{lem:correct} and Lemma~\ref{lem:updown}. 
We will now analyze the running time of the algorithm. The procedures
\Down\ and \Up\ uses time linear in the size of the current state
and the state computed.
By Lemma~\ref{lem:statesize} the size of each state is $O(l_P)$.
Each step in the depth-first traversal thus takes time $O(l_P)$,
which gives a total running time of $O(l_P n_T+n_P)$. On the other
hand consider a path $t$ in $T$. We will argue that the
computation of all the states along the path takes total time
$O(n_P+n_t)$, where $n_T$ is the number of nodes in $t$. 
To show this we need the following lemma.
\begin{lemma}\label{lem:Tpath}
Let $t$ be a path in $T$. During the computation of the states
along the path $t$, any node $x\in V(P)$ is inserted into $X^c$ at
most once.
\end{lemma}
\begin{proof}
Since $t$ is a path we only need to consider the \Down\
computations. The only way a node $x\in V(P)$ can be inserted into
$X^c$ is if $\parent(x)\in X^c$. It thus follows from
Lemma~\ref{lem:statesize} that $x$ can be inserted into $X^c$ at
most once. 
\end{proof}
It follows from Lemma~\ref{lem:Tpath} that 
the computations of the all states when $T$ is a path takes time
$O(n_P+n_T)$. Consider a path-decomposition of $T$. A
path-decomposition of $T$ is a decomposition of $T$ into disjoint
paths. We can make such a path-decomposition of the tree $T$
consisting of $l_T$ paths. Since the running time of \Up\ and
\Down\ both are linear in the size of the current and computed
state it follows from Lemma~\ref{lem:updown} that we only need to
consider the total cost of the \Down\ computations on the paths in
the path-decompostion. Thus, the algorithm uses time at most
$\sum_{t\in T}O(n_p + n_t)=O(n_P l_T+ n_T)$.

Next we consider the space used by the algorithm. Lemma~\ref{lem:statesize}
implies that $|X^c|\leq l_P$. Now consider the size of $X^p$. A
node is inserted into $X^p$ when it is removed from $X^c$. It is
removed again when inserted into $X^c$ again. Thus
Lemma~\ref{lem:Tpath} implies $|X^p| \leq n_P$ at any time. The
total space usage is thus $O(n_P+n_T)$.
To summarize we have shown,
\begin{theorem}\label{simple} For trees $P$ and $T$ the tree path subsequence
problem can be solved in $O\left(\min(l_Pn_T + n_P, n_Pl_T+n_T)\right)$ time and $O(n_P + n_T)$ space.
\end{theorem}

\section{A Worst-Case Efficient Algorithm}
In this section we consider the worst-case
complexity of TPS and present an algorithm using subquadratic
running time and linear space. The new algorithm works within our framework but does not use the $\Up$
procedure or the node dictionaries from the previous section.

Recall that using a simple linked list to represent the states we
immediately get an algorithm using $O(n_{P}n_{T})$ time and space.
We first show how to modify the traversal of $T$ and discard
states along the way such that at most $O(\log n_{T})$ states are
stored at any step in the traversal. This improves the space to
$O(n_{P}\log n_{T})$. Secondly, we decompose $P$ into small
subtrees, called \emph{micro trees}, of size $O(\log n_{T})$. Each micro tree can be represented
in a single word of memory and therefore a state uses only
$O(\ceil{\frac{n_{P}}{\log n_{T}}})$ space. In total the space used to
represent the $O(\log n_{T})$ states is $O(\ceil{\frac{n_{P}}{\log
n_{T}}} \cdot \log n_{T}) = O(n_{P} + \log n_T)$. Finally, we show how to
preprocess $P$ in linear time and space such that computing the
new state can be done in constant time per micro tree.
Intuitively, this achieves the $O(\log n_{T})$ speedup.
\subsection{Heavy Path Traversal} 
In this section we present the
modified traversal of $T$. We first partition $T$ into disjoint
paths as follows. For each node $y\in V(T)$ let $\size(y) =
|V(T(y))|$. We classify each node as either \emph{heavy} or
\emph{light} as follows. The root is light. For each internal node
$y$ we pick a child $z$ of $y$ of maximum size among the children
of $y$ and classify $z$ as heavy. The remaining children are
light. An edge to a light child is a \emph{light edge}, and an
edge to a heavy child is a \emph{heavy edge}. The heavy child of a
node $y$ is denoted $\heavy(y)$. Let $\lightdepth(y)$ denote the
number of light edges on the path from $y$ to $\roots(T)$.
\begin{lemma}[Harel and Tarjan \cite{HT1984}]\label{lightdepth}
For any tree $T$ and node $y\in V(T)$,  $\lightdepth(y) \leq \log
n_{T} + O(1)$.
\end{lemma}
Removing the light edges, $T$ is partitioned into \emph{heavy
paths}. We traverse $T$ according to the heavy paths using the
following procedure. For  node $y \in V(T)$ define:
\bigskip
\begin{relate}
  \item[$\Visit(y)$:]  
  \begin{enumerate}
  \item If $y$ is a leaf report all leaves in $X_y$ and return.
  \item Else let $y_{1}, \ldots, y_{k}$ be the light children of $y$ and let $z = \heavy(y)$.
  \item For $i:= 1$ to $k$ do:
  \begin{enumerate}
    \item Compute $X_{y_{i}} := \Down(X_{y}, y_{i})$
    \item Compute $\Visit(y_{i})$.
  \end{enumerate}
  \item Compute $X_{z} := \Down(X_{y}, z)$.
  \item Discard $X_{y}$ and compute $\Visit(z)$.
\end{enumerate}
\end{relate}
The procedure is called on the root node of $T$ with the initial
state $\{\roots(P)\}$. The traversal resembles a depth first
traversal, however, at each step the light children are visited
before the heavy child. We therefore call this a \emph{heavy path
traversal}. Furthermore, after the heavy child (and therefore all
children) has been visited we discard $X_{y}$. At any step we have
that before calling $\Visit(y)$ the state $X_{y}$ is availiable,
and therefore the procedure is correct. We have the following
property:
\begin{lemma}\label{traversal}
For any tree $T$ the heavy path traversal stores at most $\log
n_{T} + O(1)$ states.
\end{lemma}
\begin{proof}
At any node $y \in V(T)$ we store at most one state for each of
the light nodes on the path from $y$ to $\roots(T)$. Hence, by
Lemma~\ref{lightdepth} the result follows. 
\end{proof}
Using the heavy-path traversal immediately gives an $O(n_Pn_T)$ time and $O(n_P\log n_T)$ space algorithm. In the following section we improve the time and space by an additional $O(\log n_T)$ factor.

\subsection{Micro Tree Decomposition} 
In this section we present
the decomposition of $P$ into small subtrees. A \emph{micro tree}
is a connected subgraph of $P$. A set of micro trees $MS$ is a
\emph{micro tree decomposition} iff $V(P) = \cup_{M \in MS} V(M)$
and for any $M, M' \in MS$, $(V(M) \backslash \{\roots(M)\}) \cap
(V(M') \backslash \{\roots(M')\}) = \emptyset$.
 Hence, two micro trees in a decomposition share at most one node and this node must be the root in at least one of the micro trees. If $\roots(M') \in V(M)$ then $M$ is the \emph{parent} of $M'$ and $M'$ is the \emph{child} of $M$. A micro tree with no children is a \emph{leaf} and a micro tree  with no parent is a \emph{root}. Note that we may have several root micro trees since they can overlap at the node $\roots(P)$. We decompose $P$ according to the following classic result:
\begin{lemma}[Gabow and Tarjan \cite{GT1985}]\label{clustering}
For any tree $P$ and parameter $s > 1$, it is possible to build  a
micro tree decomposition $MS$ of $P$ in linear time such that
$|MS| = O(\ceil{n_P/s})$ and $|V(M)| \leq s$ for any $M \in MS$
\end{lemma}

\subsection{Implementing the Algorithm} In this section we show
how to implement the $\Down$ procedure using the micro tree
decomposition. First decompose $P$ according to
Lemma~\ref{clustering} for a parameter $s$ to be chosen later.
Hence, each micro tree has at most $s$ nodes and $|MS| =
O(\ceil{n_P/s})$. We represent the state $X$ compactly using a bit vector
for each micro tree. Specifically, for any micro tree $M$ we store
a bit vector $X_M = [b_{1}, \ldots, b_{s}]$, such that $X_M[i] = 1$
iff the $i$th node in a preorder traversal of $M$ is in $X$. If
$|V(M)| < s$ we leave the remaining values undefined. Later we
choose $s= \Theta(\log n_{T})$ such that each bit vector can be
represented in a single word.

Next we define a $\Down_{M}$ procedure on each micro tree $M\in
MS$. Due to the overlap between micro trees the $\Down_{M}$
procedure takes a bit $b$ which will be used to propagate
information between micro trees. For each micro tree $M \in MS$,
bit vector $X_M$, bit $b$, and $y\in V(T)$ define:
\begin{relate}
\item[$\Down_{M}(X_M, b, y)$:] Compute the state  $X'_M := \Child(\{x \in X_M \mid \lab(x) = \lab(y)\}) \cup \{x \in X_M \mid \lab(x) \neq \lab(y)\}$. If $b=0$, return $X_M'$, else return $X_M' \cup \{\roots(M)\}$.
\end{relate}
Later we will show how to implemenent $\Down_{M}$ in constant time
for $s = \Theta(\log n_{T})$. First we show how to use $\Down_M$ to
simulate $\Down$ on $P$. We define a recursive procedure $\Down$
which traverse the hiearchy of micro trees. For micro tree
$M$, state $X$, bit $b$, and $y \in V(T)$ define:
\begin{relate}
\item[$\Down(X,M,b,y)$:] Let $M_1, \ldots, M_k$ be the children of $M$.
\begin{enumerate}
\item Compute  $X_M := \Down_{M}(X_M, b, y)$.
\item For $i:=1$ to $k$ do:
\begin{enumerate}
\item Compute $\Down(X, M_i, b_{i}, y)$, where $b_{i} = 1$
iff

$\roots(M_i) \in X_M$.
\end{enumerate}
\end{enumerate}
\end{relate}
Intuitively, the $\Down$ procedure works in a top-down fashion
using the $b$ bit to propagate the new state of the root of micro
tree. To solve the problem within our framework we initially
construct the state representing $\{\roots(P)\}$. Then, at each
step we call $\Down(R_{j}, 0, y)$ on each root micro tree $R_{j}$.
We formally show that this is correct:
\begin{lemma}
The above algorithm correctly simulates the $\Down$ procedure on
$P$.
\end{lemma}
\begin{proof}
Let $X$ be the state and let $X' :=\Down(X, y)$. For
simplicity, assume that there is only one root micro tree $R$.
Since the root micro trees can only overlap at $\roots(P)$ it is
straightforward to generalize the result to any number of roots.
We show that if $X$ is represented by bit vectors at each micro
tree then calling $\Down(R, 0, y)$ correctly produces the new
state $X'$.

If $R$ is the only micro tree then only line 1 is executed. Since
$b = 0$ this produces the correct state by definition of
$\Down_{M}$. Otherwise, consider a micro tree $M$ with children
$M_{1}, \ldots, M_{k}$ and assume that $b = 1$ iff $\roots(M) \in
X'$. Line 1 computes and stores the new state returned by
$\Down_{M}$. If $b=0$ the correctness follows immediately. If
$b=1$ observe that $\Down_{M}$ first computes the new state and
then adds $\roots(M)$. Hence, in both cases the state of $M$ is
correctly computed. Line 2 recursively computes the new state of
the children of $M$. 
\end{proof}

If each micro tree has size at most $s$ and $\Down_{M}$ can be
computed in constant time it follows that the above algorithm
solves TPS in $O(\ceil{n_{P}/s})$ time. In the following section we show
how to do this for $s = \Theta(\log n_{T})$, while maintaining
linear space.

\subsection{Representing Micro Trees} In this section we show
how to preprocess all micro trees $M \in MS$ such that
$\Down_{M}$ can be computed in constant time. This preprocessing
may be viewed as a ``Four Russian Technique''~\cite{ADKF1970}. To
achieve this in linear space we need the following auxiliary
procedures on micro trees. For each micro tree $M$, bit vector
$X_M$, and $\alpha \in \Sigma$ define:
\begin{relate}
\item[$\Child_{M}(X_M)$:] Return the bit vector of nodes in $M$ that are children of nodes in $X_M$.
\item[$\Eq_{M}(\alpha)$:] Return the bit vector of nodes in  $M$ labeled $\alpha$.
\end{relate}
By definition it follows that:
\begin{equation*}
\begin{aligned}
\Down_{M}(X_M,b, y) &=
\begin{cases}
\Child_{M}(X_M \cap \Eq_M(\lab(y)))\; \cup \\
\quad (X_M \backslash (X_M \cap  \Eq_M(\lab(y)))      & \text{if $b = 0$}, \\
\Child_{M}(X_M \cap \Eq_M(\lab(y))) \; \cup \\
\quad (X_M \backslash (X_M \cap \Eq_M(\lab(y))) \cup \{\roots(M)\}
& \text{if $b=1$}.
\end{cases} \\
\end{aligned}
\end{equation*}
Recall that the bit vectors are represented in a single word. Hence,
given $\Child_{M}$ and $\Eq_{M}$ we can compute $\Down_M$ using
standard bit-operations in constant time.

Next we show how to efficiently implement the operations. For each
micro tree $M \in MS$ we store the value $\Eq_{M}(\alpha)$ in a
hash table indexed by $\alpha$. Since the total number of
different characters in any $M\in MS$ is at most $s$, the hash
table $\Eq_{M}$ contains at most $s$ entries. Hence, the total
number of entries in all hash tables is $O(n_{P})$. Using perfect
hashing we can thus represent $\Eq_{M}$ for all micro trees, $M\in
MS$, in $O(n_{P})$ space and $O(1)$ worst-case lookup time. The
preprocessing time is expected $O(n_{P})$ w.h.p.. To get a worst-case bound we 
use the deterministic dictionary of Hagerup et. al.
\cite{HMP2001} with $O((n_{P})\log (n_{P}))$ worst-case preprocessing
time. 

Next consider implementing $\Child_{M}$. Since this
procedure is independent of the labeling of $M$ it suffices to
precompute it for all \emph{topologically} different rooted trees
of size at most $s$. The total number of such trees is less than
$2^{2s}$ and the number of different states in each tree is at
most $2^{s}$. Therefore $\Child_{M}$ has to be computed for a
total of $2^{2s}\cdot 2^{s} = 2^{3s}$ different inputs. For any
given tree and any given state, the value of $\Child_{M}$ can be
computed and encoded in $O(s)$ time. In total we can precompute
all values of $\Child_{M}$ in $O(s2^{3s})$ time. Choosing the
largest $s$ such that $3s + \log s \leq n_{T}$ (hence $s =
\Theta(\log n_{T})$) we can precompute
all values of $\Child_{M}$ in $O(s2^{3s}) = O(n_{T})$ time and space. Each of
the inputs to $\Child_{M}$ are encoded in a single word such that
we can look them up in constant time.

Finally, note that we also need to report the leaves of a state efficiently since this is needed in line 1 in the $\Visit$-procedure. To do this compute the state $L$ corresponding to all leaves in $P$. Clearly, the leaves of a state $X$ can be computed by performing a bitwise AND of each pair of bit vectors in $L$ and $X$. Computing $L$ uses $O(n_{P})$ time and the bitwise AND operation uses $O(\ceil{n_{P}/s})$ time.

Combining the results, we decompose $P$, for $s$ as described
above, and compute all values of $\Eq_{M}$ and $\Child_{M}$. Then, we solve TPS using the heavy-path
traversal. Since $s = \Theta(\log n_{T})$, from Lemmas~\ref{traversal} and \ref{clustering} we have the following
theorem:
\begin{theorem}\label{faster} For trees $P$ and $T$ the tree path subsequence problem can be solved in $O(\frac{n_Pn_T}{\log n_T} +n_T+ n_P\log n_P)$ time and $O(n_P + n_T)$ space.
\end{theorem}
Combining the results of Theorems~\ref{simple} and \ref{faster} proves Theorem~\ref{main}.

\section{Acknowledgments}
The authors would like to thank Anna {\"O}stlin Pagh for many helpful comments.

\bibliographystyle{abbrv}
\bibliography{pub}

\end{document}